\documentclass[twocolumn,showpacs,preprintnumbers,amsmath,amssymb]{revtex4}
\usepackage{dcolumn}
\usepackage{bm}
\usepackage[dvipdf]{graphicx}
\DeclareGraphicsExtensions{.jpg,.pdf,.mps,.png,.eps,.ps,.EPS}

\def\SKm#1#2{#1\choose #2}
\def\mean#1{\left\langle #1\right\rangle}
\def\not#1{\widetilde{#1}}
\begin{document}


\title{Subgraphs in random networks}

\author{S. Itzkovitz$^{1,2}$, R. Milo$^{1,2}$, N. Kashtan$^{2,3}$, G. Ziv$^1$, U. Alon{$^{1,2}$}}
\affiliation{$^1$Department of Physics of Complex Systems,
Weizmann Institute of
Science, Rehovot, Israel 76100\\
$^2$Department of Molecular Cell Biology, Weizmann Institute of
Science, Rehovot, Israel 76100\\$^3$Department of Computer Science
and Applied Mathematics, Weizmann Institute of Science, Rehovot,
Israel 76100}

\begin{abstract}

Understanding the subgraph distribution in random networks is
important for modelling complex systems. In classic Erd\H{o}s
networks, which exhibit a Poissonian degree distribution, the
number of appearances of a  subgraph G with n nodes and g edges
scales with network size as $\mean{G}\sim{N^{n-g}}$. However, many
natural networks have a non-Poissonian degree distribution. Here
we present approximate equations for the average number of
subgraphs in an ensemble of random sparse directed networks,
characterized by an arbitrary degree sequence. We find new scaling
rules for the commonly occurring case of directed scale-free
networks, in which the outgoing degree distribution scales as
$P(k)\sim{k^{- \gamma}}$. Considering the power exponent of the
degree distribution, $\gamma$, as a control parameter, we show
that random networks exhibit transitions between three regimes. In
each regime the subgraph number of appearances follows a different
scaling law, $\mean{G}\sim{N^\alpha}$ , where $\alpha=n-g+s-1$ for
$\gamma<2$, $\alpha=n-g+s+1-\gamma$ for $2<\gamma<\gamma_c$, and
$\alpha=n-g$ for $\gamma>\gamma_c$, $s$ is the maximal outdegree
in the subgraph, and $\gamma_c=s+1$. We find that certain
subgraphs appear much more frequently than in Erd\H{o}s networks.
These results are in very good agreement with numerical
simulations. This has implications for detecting network motifs,
subgraphs that occur in natural networks significantly more than
in their randomized counterparts.
\end{abstract}
\pacs{05, 89.75}
\maketitle
\section{Introduction}
Many natural systems are described as networks of interacting
components (\cite{Strogatz 2001}-\cite{Maslov 1998}). Random
networks have been studied as models of these complex systems. The
classic model for a random network is the Erd\H{o}s model
(\cite{Erdos1959}-\cite{Bollobas1985}), in which each of the
possible edges in the network exists with probability $p$. There
exists an analytical solution to many of the properties of
Erd\H{o}s networks, such as the diameter, clustering coefficient,
component size distributions, and subgraph distributions
(\cite{Erdos1959}-\cite{Bollobas1985}). The average number of
appearances $G$ of a subgraph with n nodes and g edges in a
directed network of $N$ nodes is
\begin{eqnarray}
&&\mean{G}\!=\!\lambda{\SKm{N}{n}}p^g(1-p)^{n(n-1)-g}\!\!\sim\!\!{\lambda}N^n\left(\frac{\mean{K}}{N}\!\!\right)^g\!\!\!\!\nonumber\\&&\mean{G}\!\sim
\!\!N^{n-g} \label{eq1}
\end{eqnarray}

 assuming a fixed mean connectivity $\mean{K}=pN$.
$\lambda$ is a term of order 1 which stems from the symmetry of
each subgraph. Erd\H{o}s networks have been extensively used as
models for analyzing real networks. An excellent example is the
work of Davis, Holland and Leinhardt on subgraphs in social
networks(\cite{Holland}-\cite{Wasserman}).

\begin{figure}
\begin{center}
 \includegraphics[width = 90 mm, height = 40 mm ]{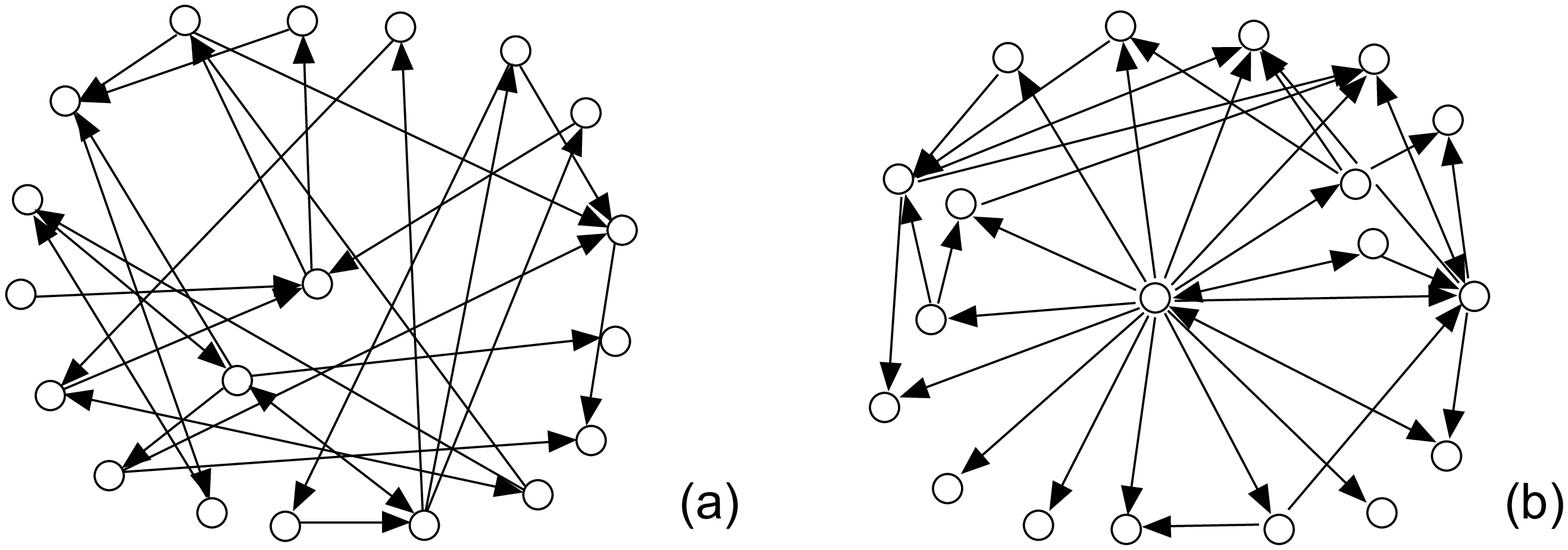}
\caption{Example of (a) Erd\"{o}s network and (b) Scale-free
network ($\gamma=2$). Mean connectivity is 1.85 in both. Notice
the hub in the scale-free network.}\label{hub1}
\end{center}
\end{figure}

Erd\H{o}s networks exhibit a Poissonian degree distribution: the
distribution of the number of edges per node is
$P(k)={\mean{k}}^ke^{-\mean{k}}/k!$. Nodes with a number of edges
much higher than the mean are exponentially rare. Many naturally
occurring networks, on the other hand, obey a long-tailed degree
sequence, often described as a power law, $P(k)\sim k^{-\gamma}$,
with $\gamma$ often between 2 and 3
(\cite{Barabasi1999}-\cite{Sole 2001}). These networks, termed
scale-free networks, are characterized by the existence of nodes
with high degree, termed hubs (Fig\ref{hub1}). The existence of
hubs dramatically influences the properties of these networks.
Some of the global properties of random networks with arbitrary
degree distribution, and specifically scale-free networks, have
been calculated. These include sizes of connected components
(\cite{Newman 2001},\cite{Aiello 2001},\cite{Molloy 1998}),
distances (\cite{Chung_diameter}), percolation thresholds
(\cite{Cohen 2002}-\cite{Newman 1999 percolation}) and clustering
coefficients (\cite{Newman book},\cite{dorogovtsev
2002a},\cite{Ravasz 2002}).

There is much current interest in the local structure of networks
(\cite{Watts 1998},\cite{Shen or 2002}-\cite{Maslov
1998},\cite{Newman book}-\cite{Eckmann2},\cite{Maslov
B}-\cite{Newman 1998}). Recently subgraph structure was analyzed
in biological and technological networks(\cite{Shen or
2002},\cite{Milo 2002}). It was found that these natural or
designed networks contain network motifs, subgraphs that occur
much more often than in an ensemble of randomized networks with
the same degree sequence. In biological networks, the network
motifs were suggested to be elementary building blocks which carry
out key information processing functions (\cite{Shen or
2002},\cite{Milo 2002}). In these studies, random networks
generation and the enumeration of their subgraphs were performed
numerically. To complement this numerical work, it would be
important to theoretically characterize the subgraph distribution
of random networks. Here we present approximate formulas for the
average number of subgraphs in an ensemble of random networks with
an arbitrary degree sequence. In the random ensemble each node has
a specified indegree, outdegree, and mutual degree. These formulas
give a very good approximation for random networks which allow for
multiple edges between nodes (more than one edge in a given
direction), as in the well-studied configuration model
(\cite{Bollobas1985},\cite{Molloy 1998}-\cite{Chung_diameter}). We
also show that they provide a reasonable approximation for
networks where multiple edges are not allowed, which represent
more realistically many naturally occurring networks. We apply
these formulas to arrive at new scaling laws for networks with a
scale-free degree distribution. We find that each subgraph has its
own scaling exponent, influenced by its topology. Considering the
power exponent of the degree distribution, $\gamma$, as a control
parameter, we show that the random networks exhibit transitions
between 3 regimes. In each regime the subgraph number of
appearances follows a different scaling law. We find that certain
subgraphs appear much more frequently than in Erd\H{o}s networks.

\section{Number of Subgraphs, approximate solution}

\begin{figure}
\includegraphics[width = 80 mm, height = 80 mm ]{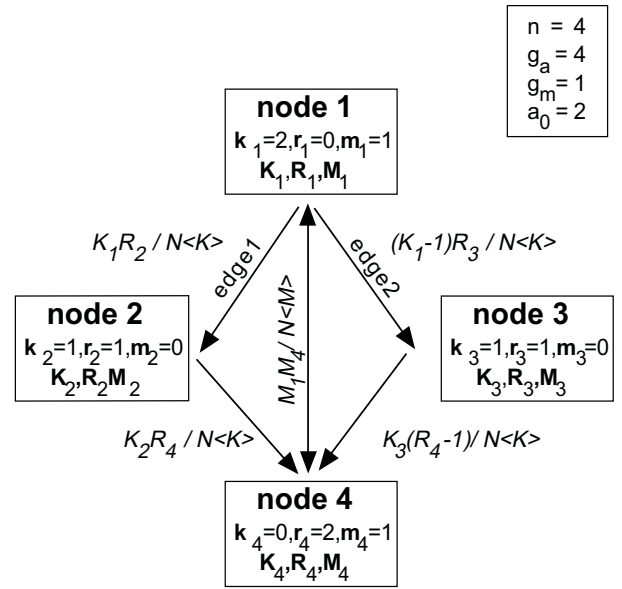}
\caption{A subgraph with one mutual edge and 4 single edges. The
subgraph degree sequences $\{k_i,r_i,m_i\}$ and node degrees
$\{K_i,R_i,M_i\}$ are displayed in bold. Edge probabilities are
displayed in plain. Using Eq. (\ref{eq6}), the mean subgraph
number of appearances in an ensemble of random networks is
$\mean{G}=2\mean{K(K-1)M}\mean{R(R-1)M}\mean{RK}^2/N\mean{K}^4\mean{M}$
}\label{explanation}
\end{figure}

The following approximation assumes sparse networks
($\mean{K}\!\ll\!{N}$). The network degree sequence is given by
the outdegree $\{K_i\}_{i=1}^N$ (the number of edges outgoing from
each node), indegree $\{R_i\}_{i=1}^N$ (the number of incoming
edges at each node), and mutual degree $\{M_i\}_{i=1}^N$ (the
number of mutual edges at each node). Mutual edges are cases where
there is a pair of edges in both directions between two nodes.
This property has been studied in social networks
(\cite{Holland}-\cite{Wasserman}) and in the world wide web
(\cite{Eckmann}). We begin by computing the probability of
obtaining an n-node subgraph with $g_a$ single edges, $g_m$ mutual
edges, subgraph outdegree sequence $\{k_j\}_{j=1}^n$, subgraph
indegree sequence $\{r_j\}_{j=1}^n$ and subgraph mutual degree
sequence $\{m_j\}_{j=1}^n$ in a given set of nodes. Consider the
example of (Fig \ref{explanation}). The probability of obtaining a
directed edge from node 1 to node 2 is approximately
\begin{eqnarray}
P(edge 1)=\frac{K_1R_2}{N\mean{K}}\label{eq2}
\end{eqnarray}\\
assuming $K_1R_2\ll{\!N\!\!\mean{K}}$ (see Appendix A). The
probability of obtaining a second edge from node 1 to node 3 is :
\begin{eqnarray}
P(edge 2|edge 1)=\frac{(K_1-1)R_3}{N\mean{K}}\label{eq3}
\end{eqnarray}\\
This reasoning applies to all the subgraph edges. The mean number
of appearances of a subgraph is found by taking the average of the
resulting expression with respect to all choices of $n$ distinct
nodes $\{\sigma_1\ldots\sigma_n\}$, and multiplying by the number
of possible choices of n nodes out of N:
\begin{eqnarray}
\mean{G}=\frac{aN^{n-g_a-g_m}}{\mean{K}^{g_a}\mean{M}^{g_m}}\mean{\prod_{j=1}^n{\SKm{K_{\sigma_j}}{k_j}}{\SKm{R_{\sigma_j}}{r_j}}{\SKm{M_{\sigma_j}}{m_j}}}_{\{\sigma\}}\label{eq4}
\end{eqnarray}\\
Where $\mean{K}$ is the average outdegree (equals the average
indegree $\mean{R}$), and $\mean{M}$ is the average mutual edge
degree. The symmetry factor $a$ is
$a_0^{-1}\prod_{j=1}^n{k_j!r_j!m_j!} $ , where $a_0$ is the number
of different permutations of the nodes that give an isomorphic
subgraph.
\begin{table*}
{\begin{tabular}{|c c|c|c|c|c|} \hline
subgraph&id&equation&transcription&neurons&www\\
\hline 6 *& \includegraphics[width = 10 mm, height = 7 mm
]{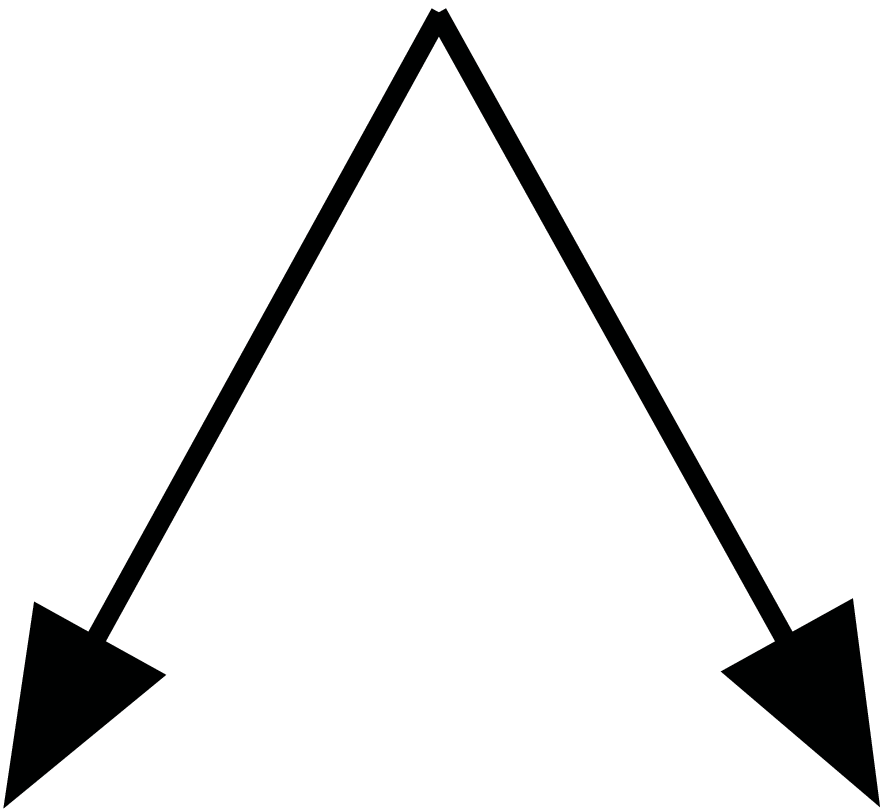}&$N{\mean{K(K-1)}}/2$&$1.2\!\!\times\!\!10^4(-0.16\%/\!\!-0.02\%)$&$4.3\!\!\times\!\!10^2(+2\%/+8\%)$&$4.7\!\!\times\!\!10^7(+0.06\%/\!\!+0.5\%)$\\\hline
 12 *& \includegraphics[width = 10 mm, height = 7 mm ]{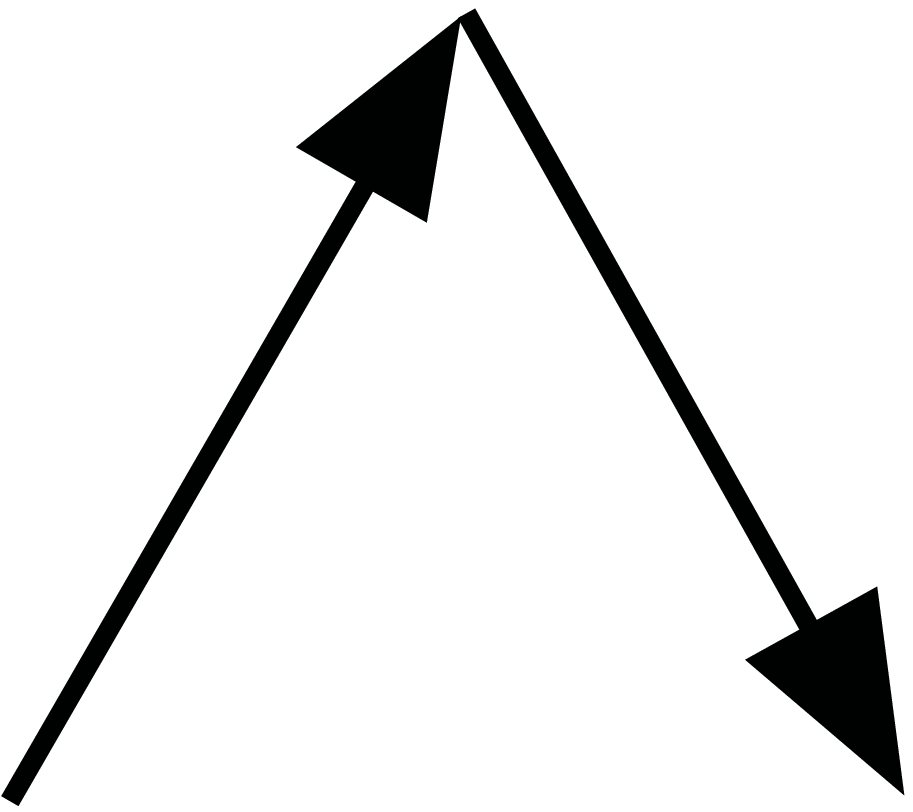}&$N\mean{KR}$&$3.6\!\!\times\!\! 10^2(+0.16\%/\!\!-0.1\%)$&$8.7\!\!\times\!\!
10^2(+2.7\%/\!\!+3.0\%)$&$2.5\!\!\times\!\!10^6(+9\%/\!\!+10\%)$\\\hline
 14 *& \includegraphics[width = 10 mm, height = 7 mm ]{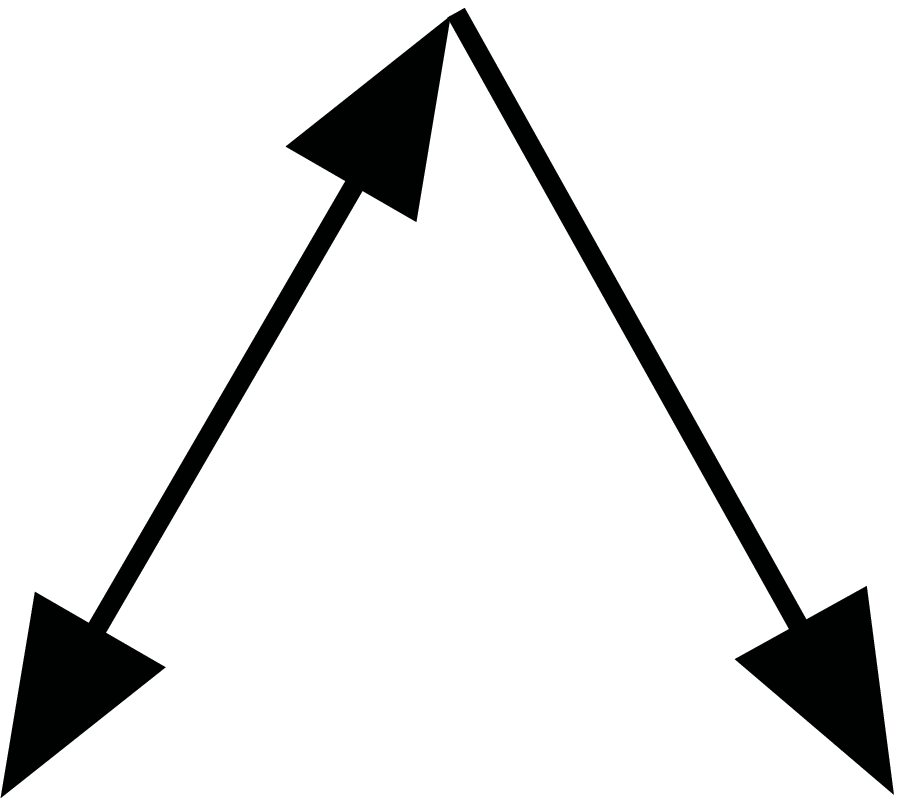}&$N\mean{KM}$&$1.9\!\!\times\!\!10^1(-0.06\%/\!\!-0.06\%)$&$8.7\!\!\times\!\!10^1(-0.15\%/\!\!+1.9\%)$&$3.8\!\!\times\!\!10^6(-0.2\%/\!\!-0.3\%)$\\\hline
 36 *& \includegraphics[width = 10 mm, height = 7 mm ]{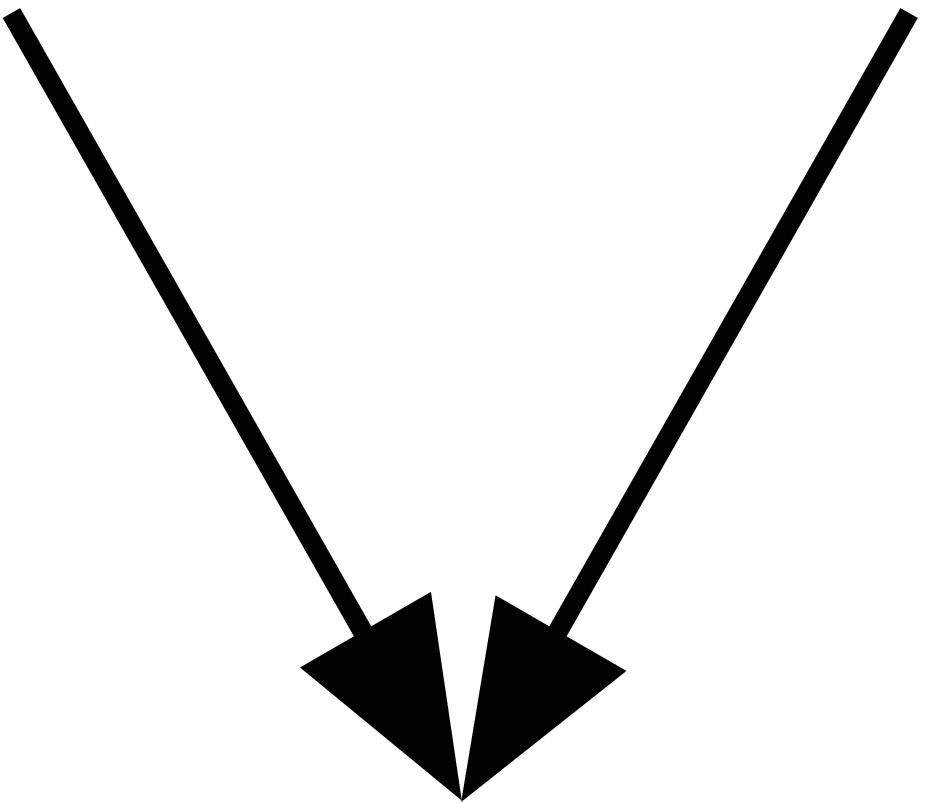}&$N\mean{R(R-1)}/2$&$9.6\!\!\times\!\!10^2(-2\%/\!\!-0.03\%)$&$6.0\!\!\times\!\!10^3(-0.4\%/\!\!+0.7\%)$&$2.2\!\!\times\!\!10^8(+0.01\%/\!\!+0.1\%)$\\
 \hline
  \hspace{-2.5mm}38& \includegraphics[width = 10 mm, height = 7 mm ]{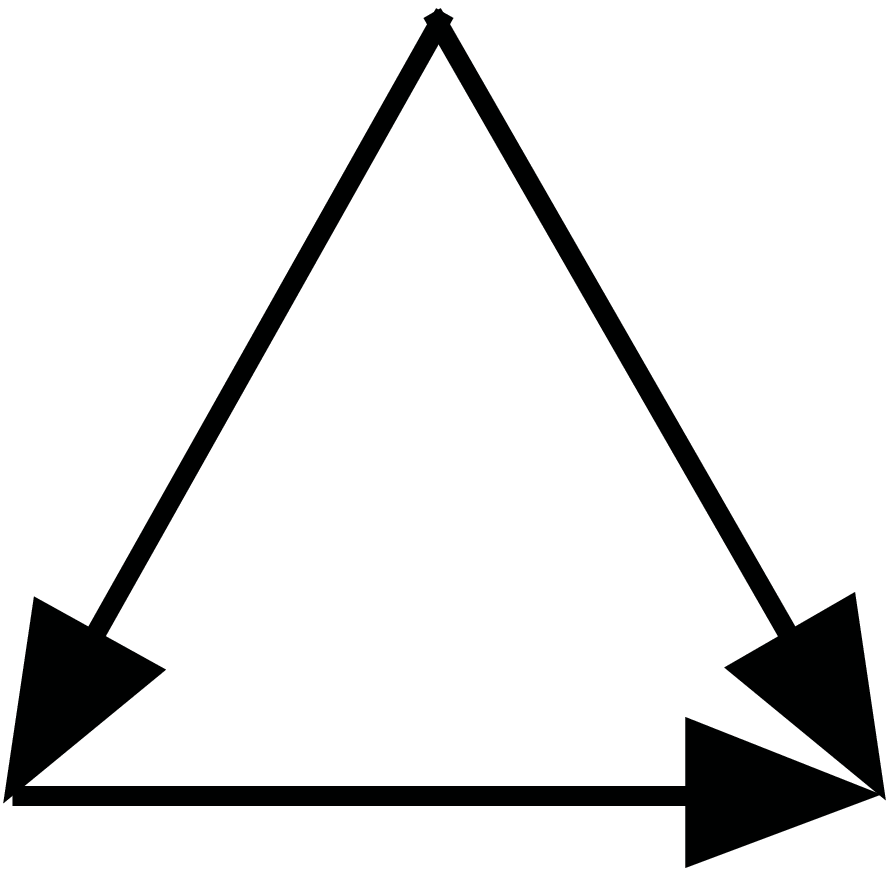}&$\!\!\mean{K(K\!\!-\!\!1)}\!\!\mean{RK}\!\!\mean{R(R-1)}\!\!/\!\!\mean{K}^3\!\!$&$1.3\!\!\times\!\!10^1(+1.6\%/\!\!+2.1\%)$&$1.2\!\!\times\!\!10^2(+0.6\%/\!\!-28\%)$&$3.4\!\!\times\!\!10^5(+0.7\%/\!\!-74\%)$\\
 \hline
  \hspace{-2.5mm}46& \includegraphics[width = 10 mm, height = 7 mm ]{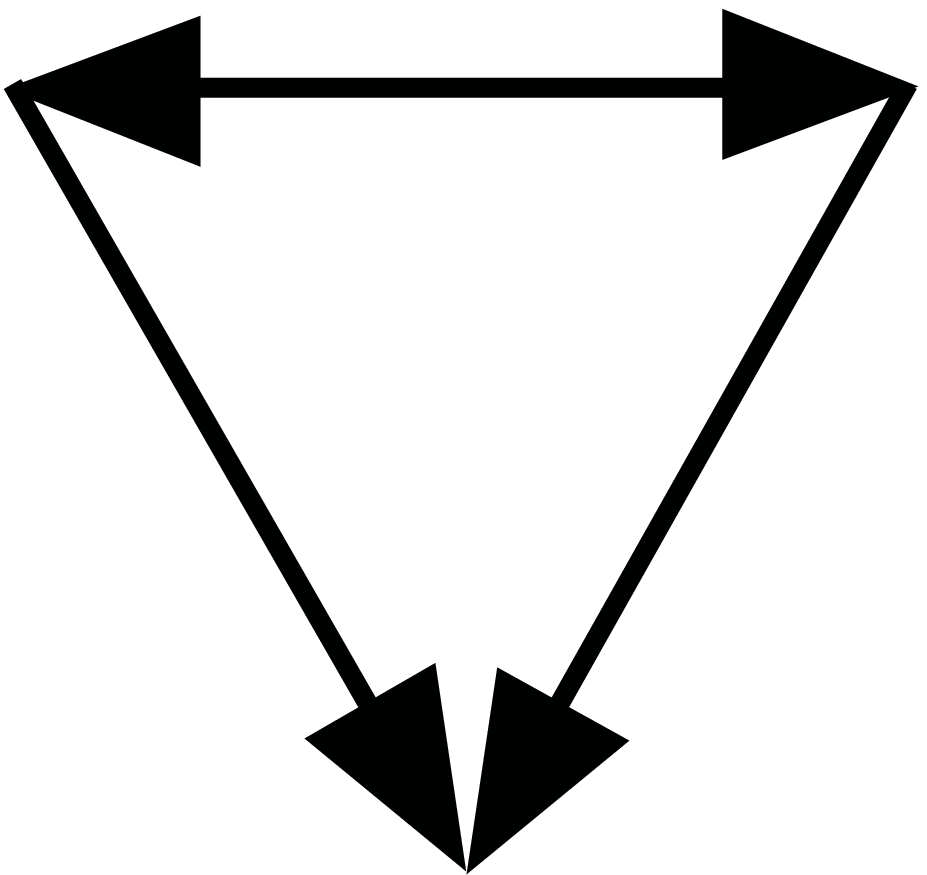}&$\!\!\mean{KM}^2\mean{R(R\!\!-\!\!1)}/{2\mean{K}^2\mean{M}}\!\!$&$0(0\%/0\%)$&$9.3\quad(-10\%/\!\!-57\%)$&$8.5\!\!\times\!\!10^3(-0.02\%/\!\!+8.8\%)$\\
 \hline
   74 *& \includegraphics[width = 10 mm, height = 7 mm ]{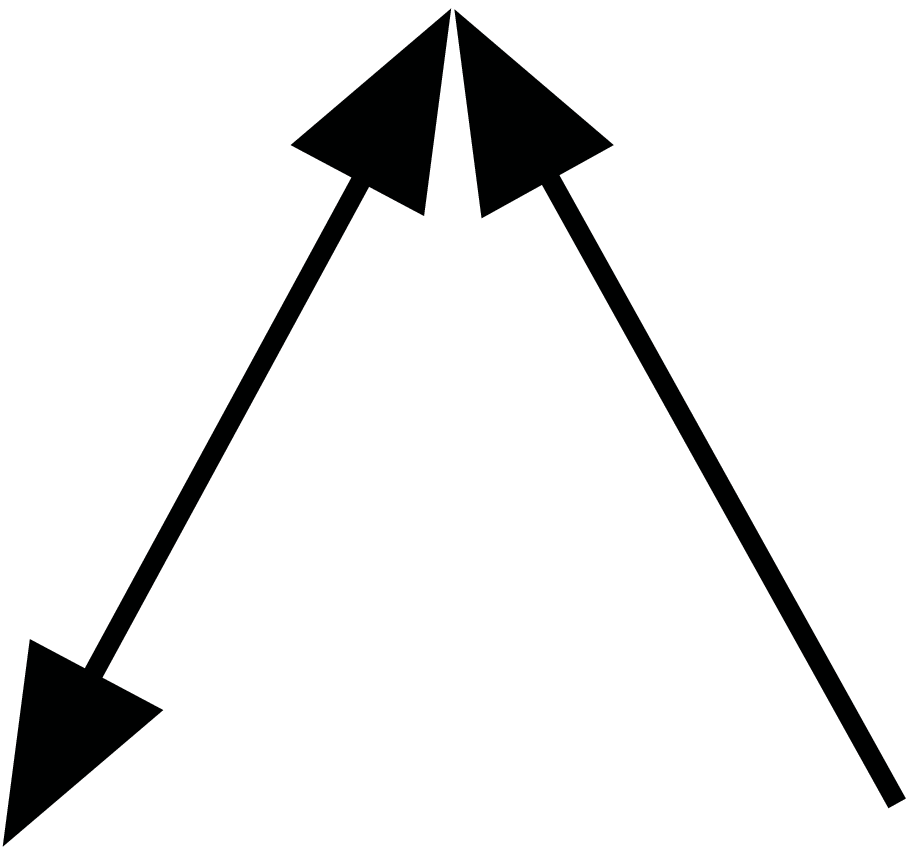}&$N\mean{RM}$&$2.9(-1.2\%/-1.8\%)$&$1.3\!\!\times\!\!10^2(+1.1\%/\!\!+1.2\%)$&$4.8\!\!\times\!\!10^6(-0.01\%/-0.01\%)$\\
 \hline
 78 *& \includegraphics[width = 10 mm, height = 7 mm ]{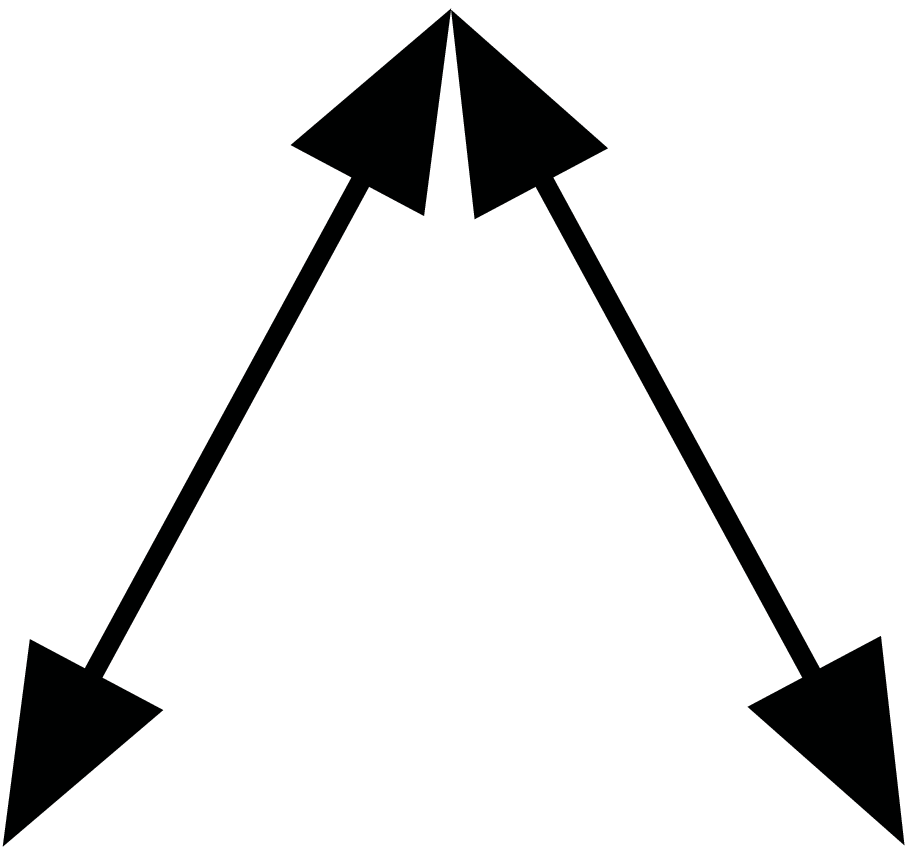}&$N\mean{M(M-1)}/2$&$0(0\%/0\%)$&$6.6\quad(-0.2\%/\!\!-0.5\%)$&$2.5\!\!\times\!\!10^7(-0.4\%/\!\!-0.4\%)$\\
 \hline
 \hspace{-2.5mm}98& \includegraphics[width = 10 mm, height = 7 mm ]{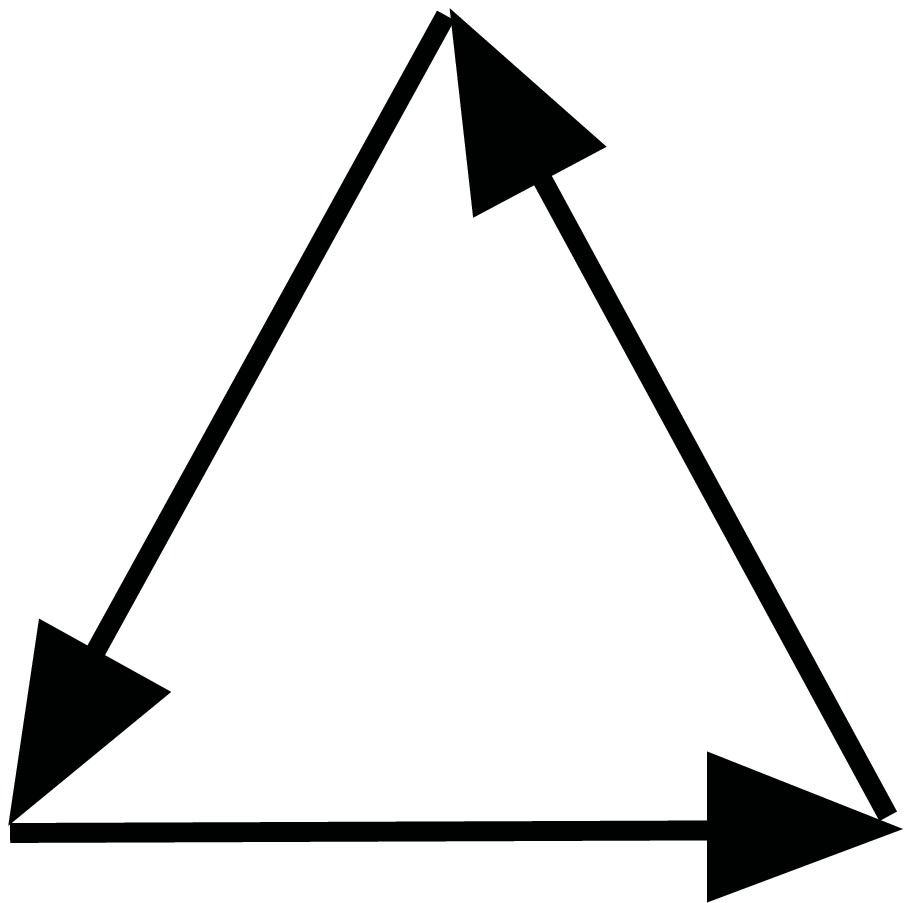}&$\mean{KR}^3/3\mean{K}^3$&$0(0\%/0\%)$&$4.5\quad(-40\%/\!\!-39\%)$&$3.3\!\!\times\!\!10^1(-31\%/\!\!-26\%)$\\
 \hline
  \hspace{-2.5mm}102& \includegraphics[width = 10 mm, height = 7 mm ]{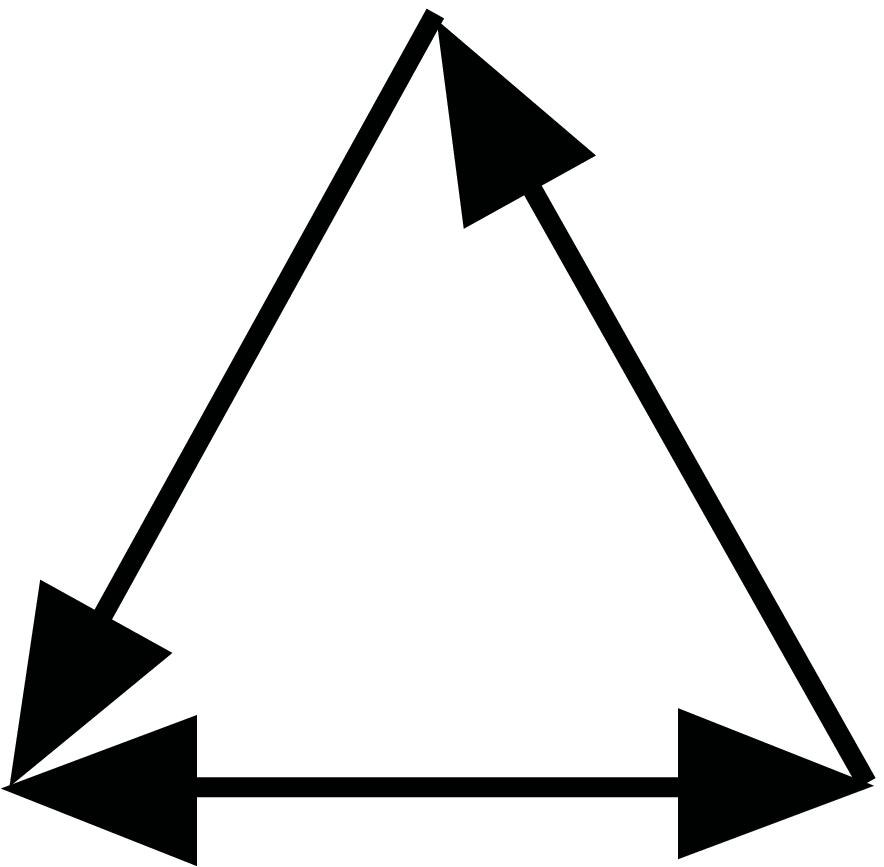}&$\!\!\mean{KM}\mean{RM}\mean{RK}/\mean{K}^2\mean{M}\!\!$&$0(0\%/0\%)$&$2\quad(-22\%/\!\!-15\%)$&$1.4\!\!\times\!\!10^2(-11\%/\!\!-4\%)$\\
 \hline
 \hspace{-2.5mm}108& \includegraphics[width = 10 mm, height = 7 mm ]{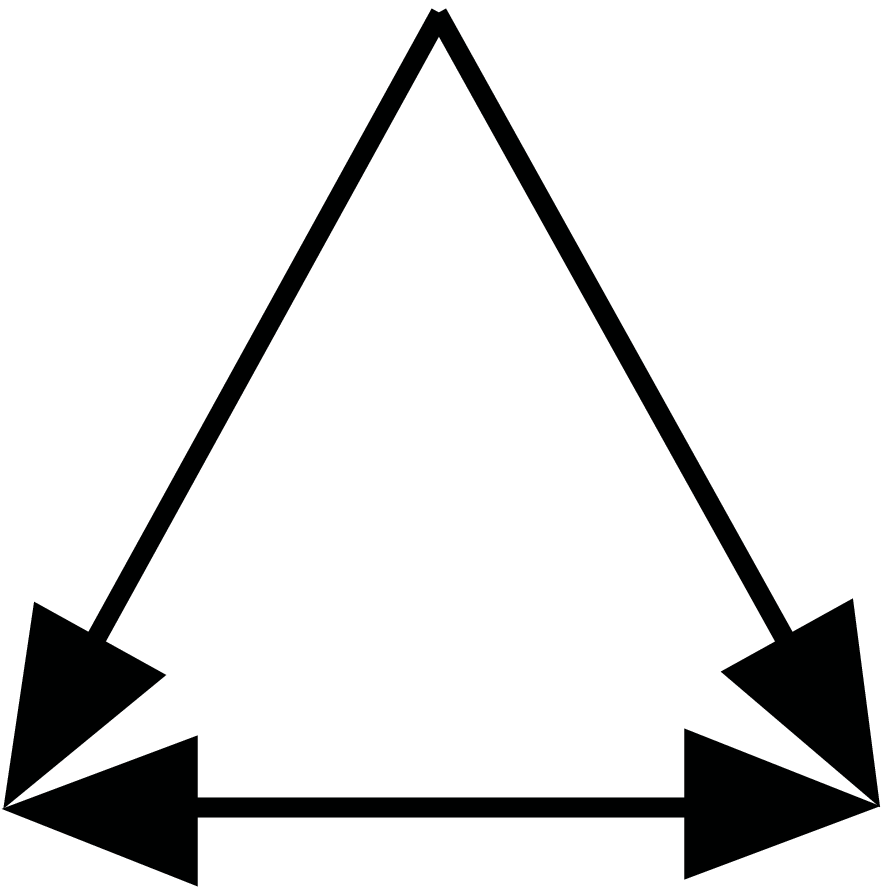}&$\!\!\mean{RM}^2\mean{K\!(\!K-1)}/2\mean{K}^2\mean{M}\!\!$&$0(0\%/0\%)$&$1.4\quad(-18\%/\!\!-6\%)$&$2.9\!\!\times\!\!10^3(-11\%/\!\!-44\%)$\\
 \hline
 \hspace{-2.5mm}110& \includegraphics[width = 10 mm, height = 7 mm ]{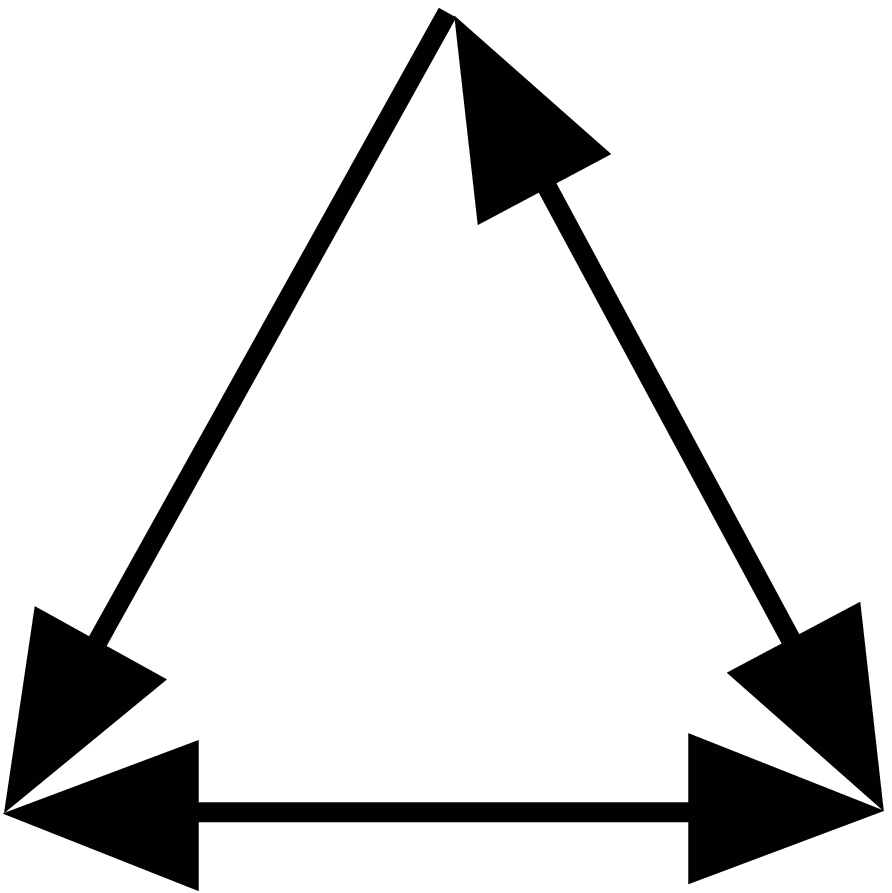}&$\mean{\!KM}\!\!\mean{\!RM}\!\!\mean{M\!(\!M\!\!-\!\!1)}\!\!/\!\!\mean{K}\!\!\mean{M}^2\!$&$0(0\%/0\%)$&$0(0\%/0\%)$&$2.3\!\!\times\!\!10^3(-1.8\%/\!\!-4\%)$\\
 \hline
  \hspace{-2.5mm}238& \includegraphics[width = 10 mm, height = 7 mm ]{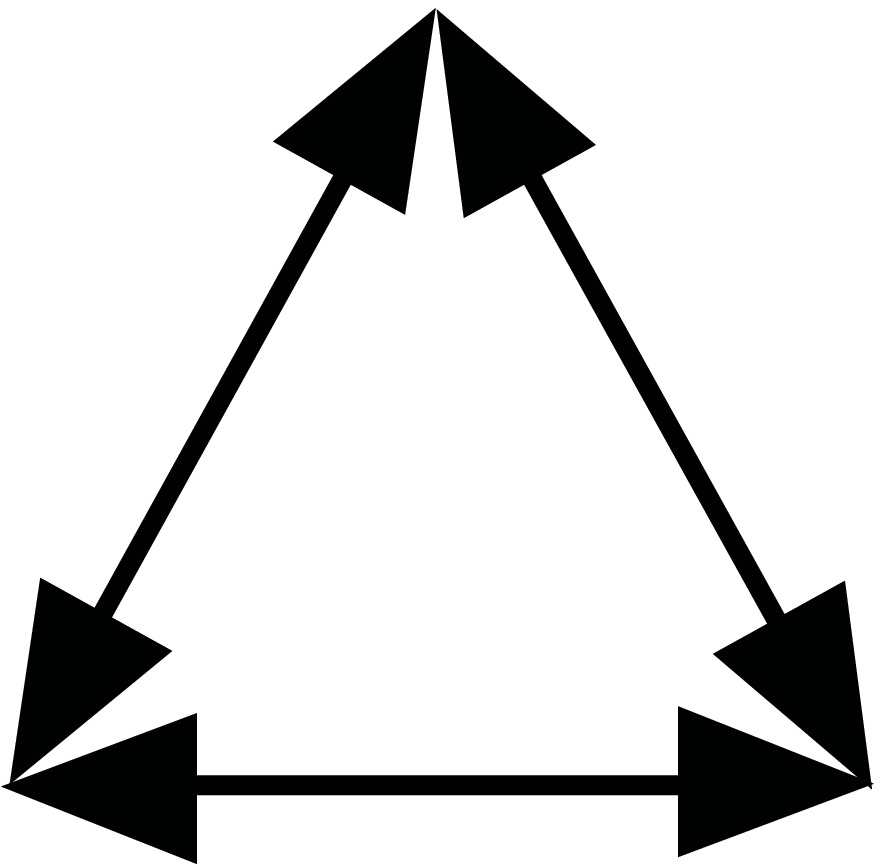}&$\mean{M(M-1)}^3/6\mean{M}^3$&$0(0\%/0\%)$&$0(0\%/0\%)$&$5\!\!\times\!\!10^4(-0.04\%/\!\!-3.6\%)$\\
 \hline
\end{tabular}
 \caption{Mean numbers of the thirteen connected directed subgraphs in an ensemble of random networks with a given degree distribution3.
 The degree distributions are those of transcription
in the yeast \textit{S. cerevisiae}(\cite{Milo 2002}), synaptic
connections between neurons in \textit{C.elegans}(\cite{White}),
and world-wide-web hyperlinks between web pages in a single
domain(\cite{Barabasi1999}).
 Shown are the theoretical values (Eq. \ref{eq6}). The values in parentheses are the percent deviations of the direct enumeration results - using the algorithms described in \cite{Milo 2002},
  where 1000 random networks with the same degree distributions as those of the real networks were generated and all
  subgraphs were counted. The left value is the percent deviation in an ensemble which allows for multiple edges, and the right value shows the deviation for an
  ensemble which does not allow multiple edges. Values below $0.5$ were rounded to zero. In subgraphs marked with *, the theoretical values shown were
  obtained using the correction of Appendix B to the table equations. Subgraph id is determined by concatenating the rows of the subgraph adjacency matrix and representing the resulting vector as a binary
  number. The id is the minimal number obtained from all the isomorphic versions of the subgraph.
  }}\label{Table1}
\end{table*}

The average (\ref{eq4}) reduces to a product of moments of
different orders of the indegree, outdegree and mutual degree
distributions:
\begin{equation}
\mean{G}=\frac{aN^{n-g_a-g_m}}{\mean{K}^{g_a}\mean{M}^{g_m}}\prod_{j=1}^n\mean{{\SKm{K_i}{k_j}}{\SKm{R_i}{r_j}}{\SKm{M_i}{m_j}}}_i\label{eq6}
\end{equation}
where the fact that each node should participate in the summation
of only one term j introduces higher order corrections which we
neglect. For example, subgraph id102 (Table I), has n=3 nodes,
$g_a=2$ single edges and $g_m=1$ mutual edge. The subgraph degree
sequences are $k_j=\{1,1,0\}$, $r_j=\{0,1,1\}$, and
$m_j=\{1,0,1\}$. Using (\ref{eq6}) we find :
\begin{equation}
\mean{G}=\mean{id102}=\frac{\mean{KM}\mean{RM}\mean{RK}}{\mean{K}^2\mean{M}}\label{eq7}
\end{equation}
The approximation (Eq. \ref{eq6}) is exact in the case of
Erd\H{o}s networks. In Erd\H{o}s networks, both indegree and
outdegree are Poisson distributed and independent, and Eq.
(\ref{eq6}) reduces to Eq. (\ref{eq1}) .\\
For non-sparse networks, a more accurate approximation takes into
account the probabilities of a non-existent edge between two nodes
(see Appendix B).\\
 We tested the equations on random networks
taken with the degree sequence of real world networks -
transcription interactions in the yeast \textit{S.
cerevisiae}(\cite{Milo 2002}), synaptic connections between
neurons in \textit{C.elegans}(\cite{White}) and world-wide-web
hyperlinks between web pages in a single
domain(\cite{Barabasi1999}). When multiple edges in the same
direction are allowed, as in the configuration model, the
equations (\ref{eq6}) are within a few percent of the numerical
simulation results (Table I). We have also simulated random
networks in which only one edge was allowed in each direction
between any two nodes. As can be seen in Table I, the equations
(\ref{eq6}) are still within a few percent of the numerical
simulation results for most subgraphs. There are some
discrepancies (most notably a factor of almost 4 for subgraph id38
in the randomized world wide web network). In addition, we find
good agreement between our approximation and numerical enumeration
of subgraphs in simulated random networks with scale-free
outdegree (Fig. 3).
\begin{figure}
\begin{center}
\includegraphics[width = 80 mm, height = 50 mm ]{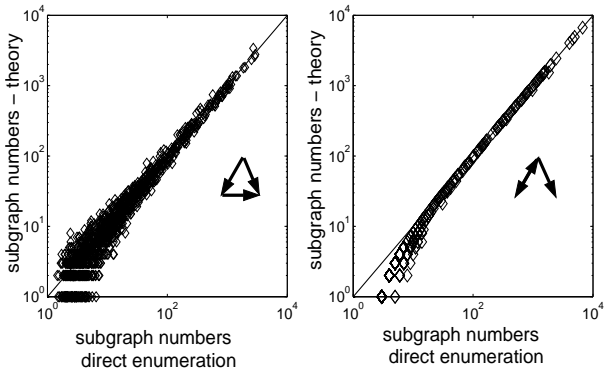}
\caption{Subgraph numbers in 1000 random networks with N=2000
nodes, with scale-free outdegree and compact indegree. The
outdegree of each node, $K_i$ was picked  from the distribution
(\ref{eq8}) , with $\gamma=2$. The networks were constructed using
the algorithm of Newman, Strogatz and Watts\cite{Newman 2001}
modified so that only a single edge in a given direction is
allowed between any two nodes. Theoretical number of appearance
were computed using the degree sequences of each network
(equations in Table I).}\label{emp_teo_ffl}
\end{center}
\end{figure}

\section{Scale-free Networks}
Scale-free networks have degree distributions that follow
$P(k)\sim{k^{-\gamma}}$ at large $k$
(\cite{Barabasi1999}-\cite{Sole 2001}). We consider directed
networks in which the outgoing edge degree is scale-free, while
the incoming edge degree distribution is Poissonian. Our results
can be easily extended to scale-free indegree. For simplicity we
choose the following form for the outgoing degree distribution for
a network with N nodes (this function was used in \cite{Newman
2001} to fit world-wide web data) :

\begin{eqnarray}
P(k)=\frac{\gamma-1}{k_0^{1-\gamma}}(k+k_0)^{-\gamma} \  \ \ \
  k<N\label{eq8}
\end{eqnarray}\\
The mean connectivity $\mean{K}$ is determined by $k_0$.

\begin{figure}
\begin{center}
\includegraphics[width = 90 mm, height = 80 mm ]{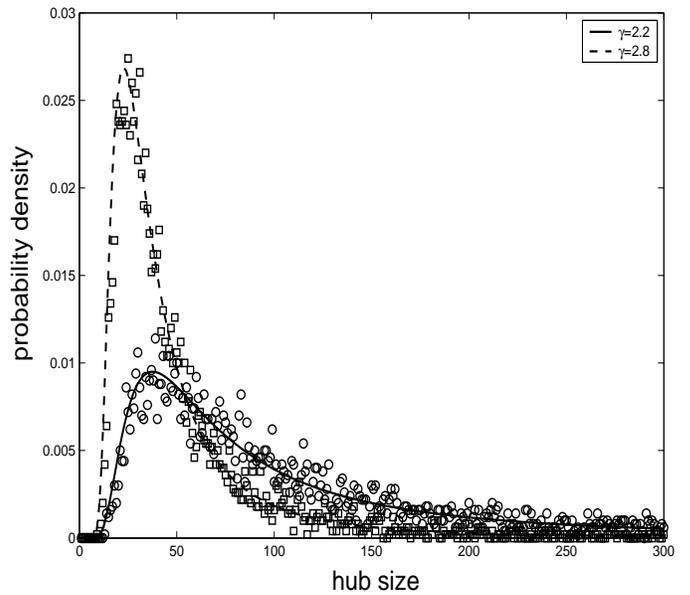}
\caption{Simulated and theoretical hub distribution for networks
with N=3000 nodes, $\gamma=2.2$ (O) or $\gamma=2.8$ ($\Box$), and
mean connectivity $\mean{K}=1.2$. Lines : theoretical
calculations, Eq. (\ref{eq9}).}\label{hub_distribution}
\end{center}
\end{figure}

The hub is the node with the maximal number of outgoing edges, T.
The hub size distribution (Fig. \ref{hub_distribution})  is :
\begin{eqnarray}
&&P(T)=NP(k=T)[P(k\leq{T})]^{N-1}=\nonumber\\&&\frac{N(\gamma-1)}{k_0}{(T/k_0)}^{-\gamma}{\left(1-{(T/k_0)}^{-\gamma+1}\right)}^{N-1}
\label{eq9}
\end{eqnarray}\\assuming $T\gg{k_0}$.
For $2<\gamma<3$, the mean hub scales as:
\begin{eqnarray}
\mean{T}=\int _1^{N-1}{{T}P(T)}dT \sim{N^{\frac{1}{\gamma-1}}}
\label{eq10}
\end{eqnarray}
where the mean is over an ensemble of random networks with the
same $\gamma$ and mean connectivity (see also
\cite{dorogovtsev_mesoscopic},\cite{Cohen_resilience} for an
alternative method of deriving this result).
At $\gamma\leq{2}$, there is a condensation effect \cite{Bianconi
2001}, where a finite fraction of the nodes have outdegree $\leq
1$, and the mean hub size becomes proportional to N. Using
(\ref{eq6}),  and assuming a compact distribution for the number
of mutual edges, we find that the subgraph distribution is
dominated by the hubs, and that the dominant term is that of the
subgraph node with maximal outdegree, $s$.   The number of
appearances of each subgraph can be shown to scale as :
\begin{eqnarray}
\mean{G}\sim{aN^{n-g-1}\mean{K}^{g-s}\sum_{i=1}^N{\SKm{K_i}{s}}\sim{N^\alpha}}
\label{eq11}
\end{eqnarray}\\
where $g=g_a+2g_m$ is the total number of edges in the subgraph
\cite{Footnote1}. We derive the scaling exponent $\alpha$ in the
following section.

\section{Transitions at different \large{$\gamma$}}
\begin{table}
\begin{tabular}{|c c| c | c |c|c|c|c|c|}
\hline
subgraph&id&n&g&s&$\alpha_{erdos}$&$\alpha_{sf}$ &$\alpha_{cond}$&$\gamma_c$\\
&&nodes&edges&&$\gamma>\gamma_c$&$2<\gamma<\gamma_c$&$\gamma\leq2$&\\
\hline
6& \includegraphics[width = 10 mm, height = 7  mm ]{id6}&3&2&2&1&$4-\gamma$&2&3\\
\hline
12&  \includegraphics[width = 10 mm, height = 7  mm ]{id12}&3&2&1&1&1&1&2\\
\hline
14& \includegraphics[width = 10 mm, height = 7  mm ]{id14}&3&3&2&0&$3-\gamma$&1&3\\
\hline
36& \includegraphics[width = 10 mm, height = 7  mm ]{id36}&3&2&1&1&1&1&2\\
\hline
38& \includegraphics[width = 10 mm, height = 7  mm ]{id38}&3&3&2&0&$3-\gamma$&1&3\\
\hline
46& \includegraphics[width = 10 mm, height = 7  mm ]{id46}&3&4&2&-1&$2-\gamma$&0&3\\
\hline
74& \includegraphics[width = 10 mm, height = 7  mm ]{id74}&3&3&1&0&0&0&2\\
\hline
78& \includegraphics[width = 10 mm, height = 7  mm ]{id78}&3&4&2&-1&$2-\gamma$&0&3\\
\hline
98& \includegraphics[width = 10 mm, height = 7  mm ]{id98}&3&3&1&0&0&0&2\\
\hline
102& \includegraphics[width = 10 mm, height = 7  mm ]{id102}&3&4&2&-1&$2-\gamma$&0&3\\
\hline
108& \includegraphics[width = 10 mm, height = 7  mm ]{id108}&3&4&2&-1&$2-\gamma$&0&3\\
\hline
110& \includegraphics[width = 10 mm, height = 7  mm ]{id110}&3&5&2&-2&$1-\gamma$&-1&3\\
\hline
238& \includegraphics[width = 10 mm, height = 7  mm ]{id238}&3&6&2&-3&$-\gamma$&-2&3\\
\hline
\end{tabular}
\\
\begin{tabular}{|c c| c | c |c|c|c|c|c|}
\hline
subgraph&id&n&g&s&$\alpha_{erdos}$&$\alpha_{sf}$ &$\alpha_{cond}$&$\gamma_c$\\
&&nodes&edges&&$\gamma>\gamma_c$&$2<\gamma<\gamma_c$&$\gamma\leq2$&\\
\hline
14&\includegraphics[width = 10 mm, height = 7 mm ]{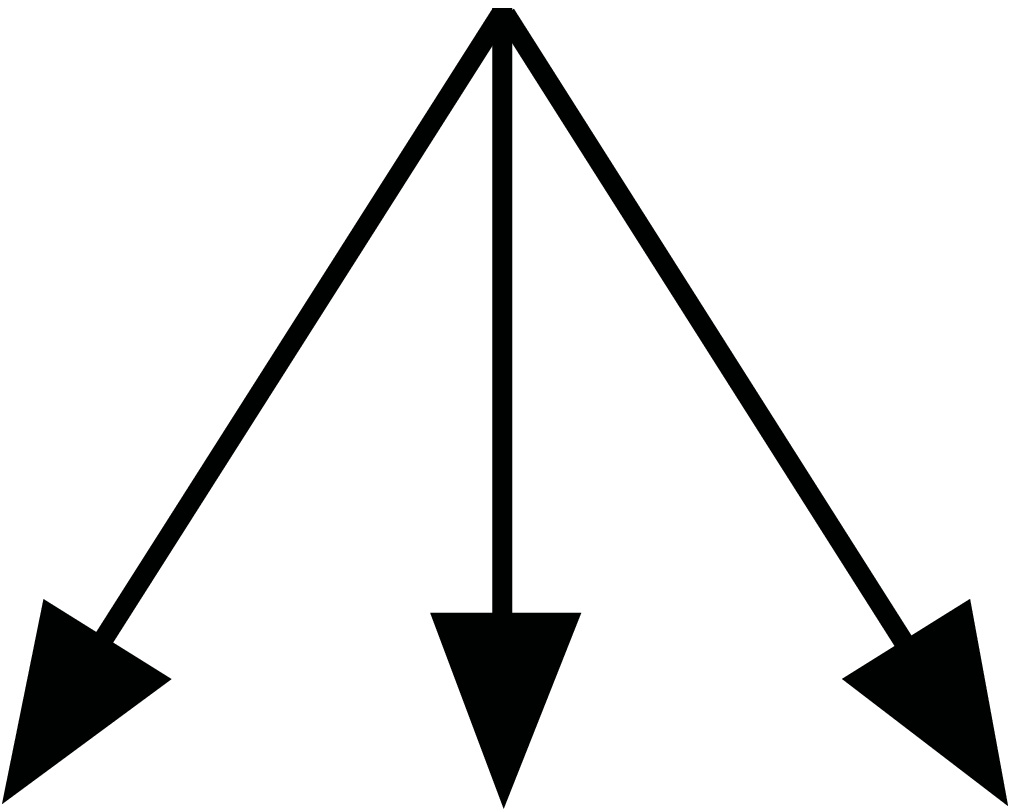}&4&3&3&1&$5-\gamma$&3&4\\
\hline
204& \includegraphics[width = 10 mm, height = 7 mm ]{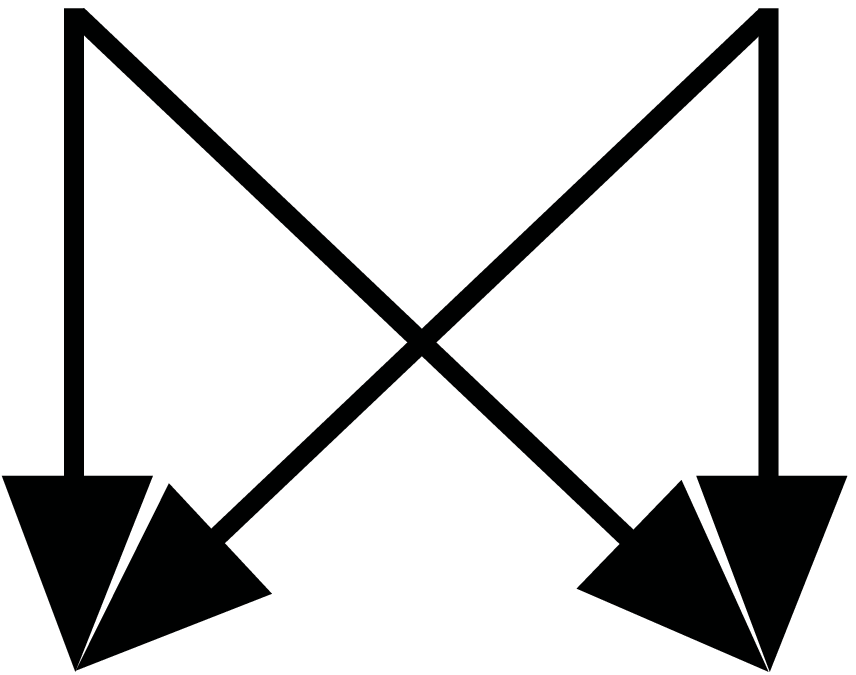}&4&4&2&0&$3-\gamma$&1&3\\
\hline
206& \includegraphics[width = 10 mm, height = 7 mm ]{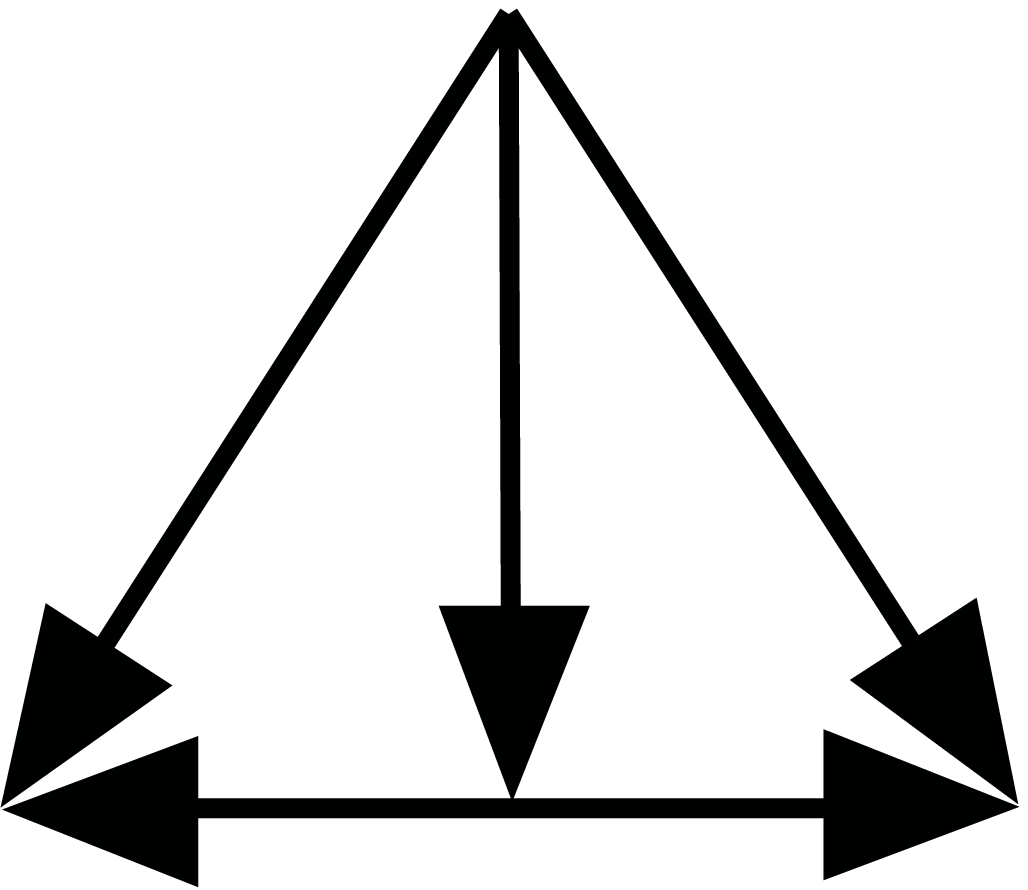}&4&5&3&-1&$3-\gamma$&1&4\\
\hline
2190& \includegraphics[width = 10 mm, height = 7 mm ]{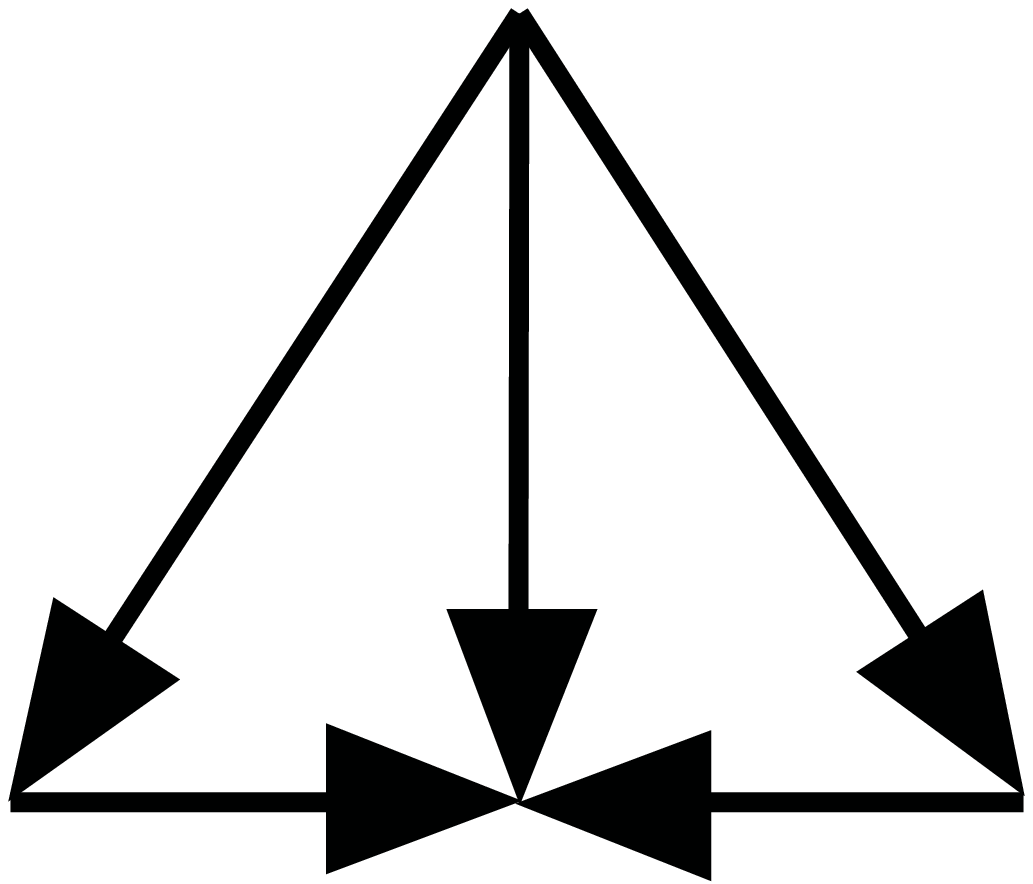}&4&5&3&-1&$3-\gamma$&1&4\\
\hline
\end{tabular}
\caption{the scaling exponent $\alpha$ of subgraph numbers for
random scale-free networks with outgoing degree exponent $\gamma$.
The subgraph numbers scale as $\mean{G}\sim{N^\alpha}$. Shown are
all thirteen 3-node connected directed subgraphs and 4 examples of
4 node subgraphs. n is the number of nodes in the subgraph, g, the
number of edges and s, the maximal degree within the subgraph. The
exponent $\alpha$ has 3 regimes : $\alpha_{erdos}$ in the
"Erd\H{o}s regime", when $\gamma>\gamma_c$, $\alpha_{sf}$ in the
"scale-free regime", when $2<\gamma<\gamma_c$, and $\alpha_{cond}$
in the "condensed regime", when $\gamma\leq2$.}\label{Table2}
\end{table}
\begin{figure}
\begin{center}
\includegraphics[width = 80 mm, height = 70 mm ]{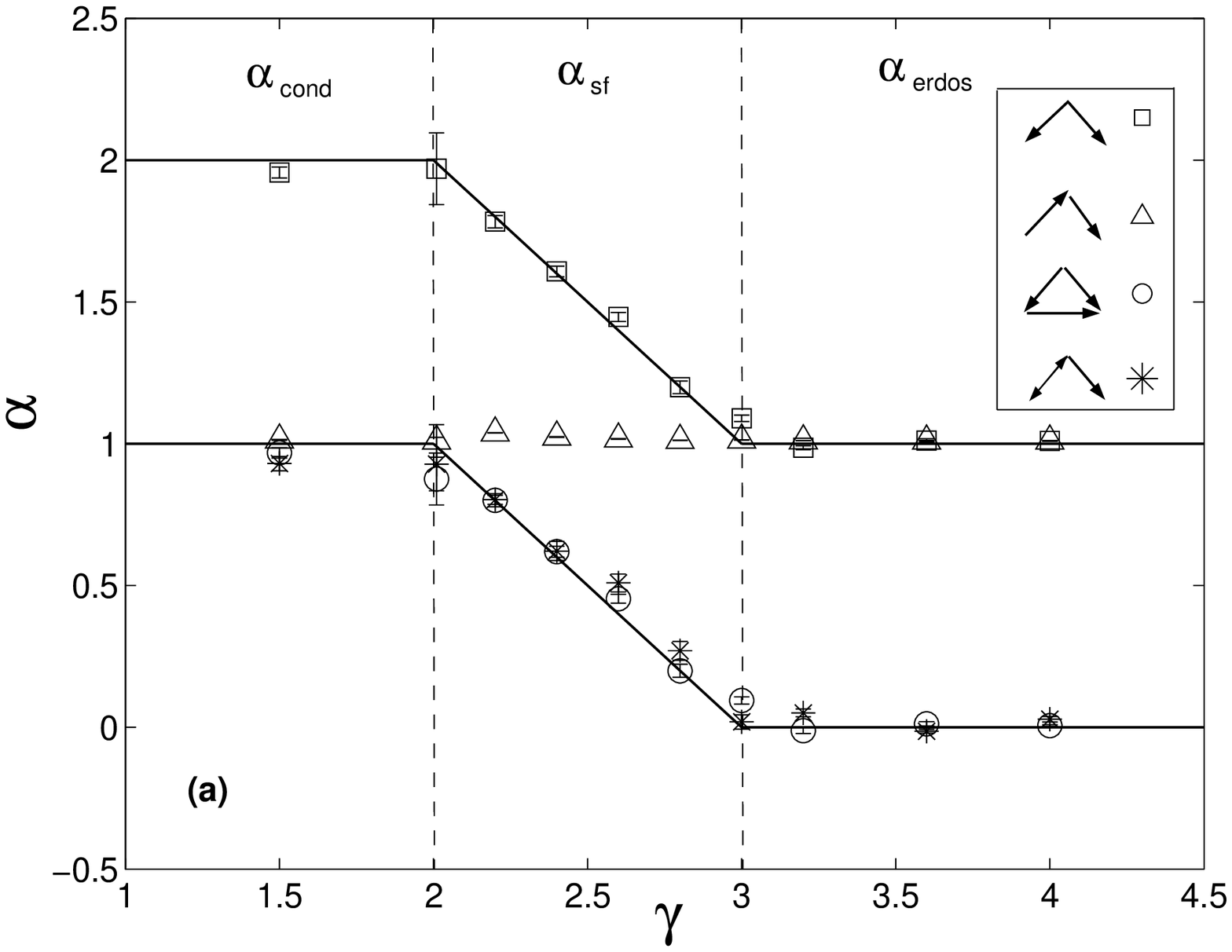}
\includegraphics[width = 80 mm, height = 70 mm ]{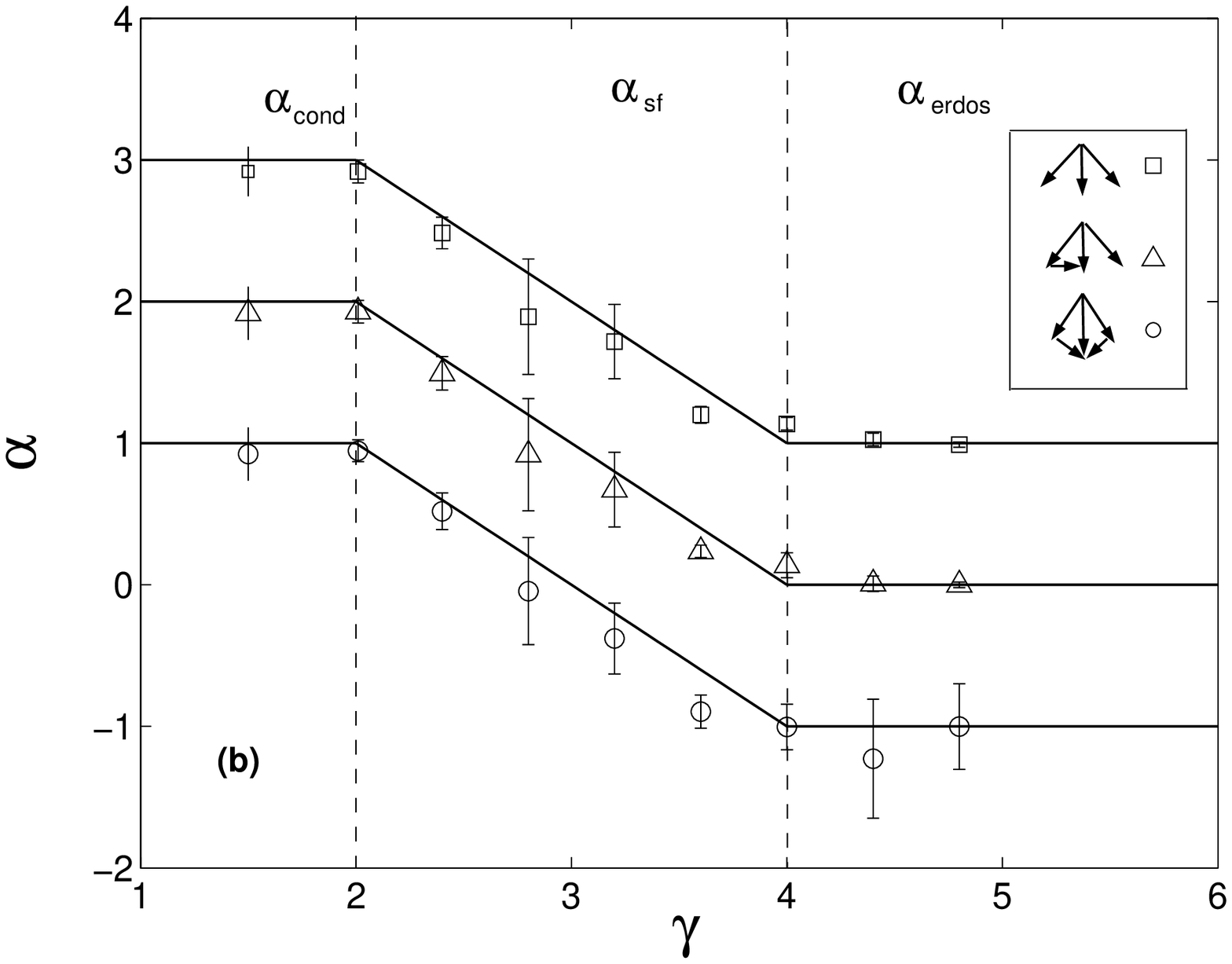}
\caption{Scaling exponent of 3-node subgraphs (a) and 4-node
subgraphs (b) as a function of $\gamma$. The exponent $\alpha$ was
obtained from the slope of a log-log fit of the number of
subgraphs vs. network size, for 9 different network sizes
(30,100,300,500,1000,1500,2000,2500,3000) averaged over 5000
randomized networks for each size and outdegree powerlaw $\gamma$.
All the networks had mean connectivity $\mean{K}=1.2$. The
exponent $\alpha$ displays three regimes, $\gamma<2$ (the
condensed regime), $2<\gamma<\gamma_c$ (the scale-free regime),
$\gamma>\gamma_c$ (Erd\H{o}s regime). }\label{scaling}
\end{center}
\end{figure}

The subgraph numbers scale as
\begin{equation}
\mean{G}\sim{N^\alpha}
\end{equation}
We find three different regimes, in each of which the scaling
exponent $\alpha$ behaves differently. Taking an ensemble average
by integrating the largest term in Eq. (\ref{eq11}) over the hub
distribution (\ref{eq9}) we get:
\begin{equation}
\mean{G}\sim{N^{n-g-1}\int_1^{N-1}T^sP(T)dT}\label{eq12}
\end{equation}
For $\gamma\leq2$ the network is in a condensed regime, where the
hub $T=O(N)$. In this regime :
\begin{eqnarray}
\mean{G}\sim{N^{n-g+s-1}}\label{eq13}
\end{eqnarray}
For $2<\gamma<\gamma_c$ substituting Eq. \ref{eq9} in Eq.
\ref{eq12} yields :
\begin{eqnarray}
\mean{G}\sim{N^{n-g+s-\gamma+1}}\label{eq13b}
\end{eqnarray}
In this regime, the tail of P(T) is the dominant contribution to
the integral. Finally at values above a critical $\gamma$, another
transition occurs, where $\alpha$ equals the scaling exponent in
Erd\H{o}s networks, $\alpha=n-g=\alpha_{erdos}$. The critical
$\gamma$ is $\gamma_c$:
\begin{eqnarray}
{\gamma}_c=s+1\label{eq14}
\end{eqnarray}
In this regime, the hubs no longer contribute significantly to the
subgraph distribution. In summary, $\mean{G}\sim{N^\alpha}$, where
$\alpha$ is :

\begin{eqnarray}
\alpha=\left\{\begin{matrix}n-g+s-1&\gamma\leq{2}\cr
n-g+s-\gamma+1&\quad2<\gamma<s+1\cr n-g&\quad\quad\gamma\geq{s+1}
\end{matrix} \right.
\end{eqnarray}\\
Table \ref{Table2} shows the expected scaling exponent for the 13
connected directed 3-node subgraphs, as well as for several 4-node
subgraphs. The scaling laws agree very well with numerical results
(Fig. 5). The three regimes of scaling are clearly seen. Note that
the topology of each subgraph effects its scaling, through the
subgraph maximal outdegree, $s$. These results can be easily
extended to the case of scale-free indegree and non-directed
networks. For loops of any size in non-directed networks the
critical ${\gamma}$ is ${\gamma}_c=3$. At $\gamma>3$, loop numbers
scale as $N^0$. This is consistent with \cite{bianconi_hloops},
which showed logarithmic corrections for the number of loops in
Barabasi-Albert scale-free networks, which have $\gamma=3$.


\section{Discussion}
To summarize, we have presented an approximate solution for the
average number of directed connected subgraphs in an ensemble of
random networks with arbitrary degree sequence. We have presented
scaling formulas for the number of subgraphs in scale-free random
networks, and showed that the subgraph numbers can be very
different from those in Erd\H{o}s random networks. Whereas in
Erd\H{o}s random networks the scaling exponent is strictly
determined by the number of nodes and edges of the subgraph, in
scale-free random networks, the exact topology of the subgraph
determines the scaling exponent. We showed that the scaling
exponent $\alpha$ exhibits three different scaling laws in three
regimes, depending on the control parameter $\gamma$ (the power of
the degree distribution). In the common case of scale-free
networks with $\gamma$ between 2 and 3, there are many more
subgraphs which contain a node connected to more than one other
node than in the corresponding Erd\H{o}s networks with the same
mean connectivity. For example, the feed-forward loop, (id38 in
Table I) is much more common for $\gamma<3$. At $\gamma=2.5$, the
number of feed-forward-loops scales as $N^{0.5}$, as opposed to
$N^0$ in Erd\H{o}s networks. On the other hand, subgraphs such as
the 3-node cycle (id98 in Table I) have the same scaling, $N^0$,
as in Erd\H{o}s networks.

This study adds to our understanding of the random network models
to which real-world networks are compared. It highlights the
importance of using random networks that preserve the single and
mutual degree sequence of the real network. Our approach may be
readily extended to networks with multiple colors of edges. The
present results may be useful for enumerating subgraphs in very
large random networks which are beyond the reach of current numerical algorithms. 

\appendix
\section{Edge Probabilities}
Here we give a more detailed derivation for the edge probabilities
used in Eq. (\ref{eq2},\ref{eq3}). Without loss of generality we
treat a network with no mutual edges. We denote by $E=N\mean{K}$
the total number of edges. We begin by calculating the probability
that no edge connects a source node with $K$ outgoing edges and a
target node with $R$ incoming edges. This happens when all $K$
edges connect to a set of nodes $\{\sigma_i\}_{i=1}^{k}$ which
does not contain the target node:
\begin{equation}
p(no\
edge|\{\sigma_i\})=\prod_{k=0}^{K-1}{\left(1-\frac{R}{E-R'-\sum_{i=1}^{k}R_{\sigma_i}}\right)}
\end{equation}\\
where $R'$ is the indegree of the source node (we do not allow
self edges). The probability of having no edge is obtained by
summing over all possible sets $\{\sigma_i\}_{i=1}^{k}$ :\\
\begin{equation}
p(no\
edge)=\frac{1}{\small{K!}{\SKm{N-2}{K}}}\sum_{\{\sigma\}}{\prod_{k=0}^{K-1}{\left(1-\frac{R}{E-R'-\sum_{i=1}^{k}R_{\sigma_i}}\right)}}
\end{equation}\\
Assuming $\max{\sum_{i=1}^{k}R_{\sigma_i}}\!\ll\!{E}$, and taking
the complement as the probability of an edge existing, we obtain:
\begin{equation}
p(edge)=1-(1-\frac{R}{N\mean{K}})^K=1\!-\!e^{-K\!R/{N\mean{K}}}\sim\frac{KR}{N\mean{K}}
\end{equation}
where our last approximation assumes $KR\!\!\ll{\!\!N\!\mean{K}}$.
Intuitively, this result can be understood as $K$ attempts for the
source node to connect to the target node with a probability of
$R/N\!\!\mean{K}$ at each attempt. $R/N\!\!\mean{K}$ is the
probability of an arbitrary edge connecting into the target node.
Pairs of nodes in which $KR$ is of the order of $N\!\mean{K}$ will
contribute multiple edges in the same direction in the
approximation, leading to over-estimation of subgraph numbers in
the simulated networks where multiple edges are not allowed (Table
I).
\section{Non-sparse networks}
In calculating the number of appearances of subgraphs in
non-sparse networks, a more accurate approximation takes into
account the probabilities of a non-existent edge between two
nodes. For such subgraphs, in addition to the specified subgraph,
Eq. (\ref{eq6}) counts  a set of subgraphs, with the null edges
replaced by single or mutual edges. The corrections for the 3-node
subgraphs are :
\begin{eqnarray}
&&\mean{id6*}=\mean{id6}-\mean{id38}-\mean{id108}\nonumber
\\
&&\mean{id12*}=\mean{id12}-\mean{id38}-\mean{id102}\nonumber\\
&&\mean{id14*}=\mean{id14}-\mean{id46}-\mean{id102}-\mean{id110}\nonumber\\
&&\mean{id36*}=\mean{id36}-\mean{id38}-\mean{id46}\nonumber\\
&&\mean{id74*}=\mean{id74}-\mean{id102}-\mean{id108}-\mean{id110}\nonumber\\
&&\mean{id78*}=\mean{id78}-\mean{id110}-\mean{id238}\nonumber\\
\label{eq7B}
\end{eqnarray}
where $\mean{G}$ represents the values obtained from Eq.
(\ref{eq6}), and $\mean{G*}$ is the corrected value. Generally,
for larger subgraphs the corrections made will be of an
inclusion-exclusion type.

\section{Subgraph Enumeration}
In numerically enumerating the subgraphs we combine a dynamic
programming method(\cite{Milo 2002}), which is applied generally
for n-node subgraphs with $n\geq4$, and a more rapid calculation,
based on adjacency matrix operations, used for 3-node subgraphs.
The method generalizes the results of (\cite{Harari}). Here we
give formulas for the thirteen 3-node connected directed subgraphs
based on the adjacency matrix. The network adjacency matrix is
denoted by $M$, where $M_{ij}=1$ if a directed edge exists from
node i to node j. We begin by dividing the network into a network
containing only antisymmetric arrows, whose adjacency matrix will
be denoted by $A$, and a network containing only mutual arrows,
whose symmetric adjacency matrix will be denoted as $S$ .
\begin{eqnarray}
M=A+S\label{eq16}
\end{eqnarray}
\begin{table}
\begin{center}
\begin{tabular}{|c c|c|}
\hline
subgraph&id&formula\\
\hline
6 & \includegraphics[width = 6 mm, height = 5 mm ]{id6}& $(\sum{A'A\cdot\not{M}\cdot\not{M'}}-tr{A'A})/2$\\
\hline
12 &\includegraphics[width = 6 mm, height = 5 mm ]{id12}&$\sum{A^2\cdot\not{M}\cdot\not{M'}}$\\
\hline
14 &\includegraphics[width = 6 mm, height = 5 mm ]{id14}&$\sum{SA\cdot\not{M}\cdot\not{M'}}$\\
\hline
36& \includegraphics[width = 6 mm, height = 5 mm ]{id36}&$(\sum{AA'\cdot\not{M}\cdot\not{M'}}-tr{AA'})/2$\\
\hline
38&\includegraphics[width = 6 mm, height = 5 mm ]{id38}&$\sum{A^2\cdot{A}}$ \\
\hline
46&\includegraphics[width = 6 mm, height = 5 mm ]{id46}&$(\sum{AA'\cdot{S}})/2$ \\
\hline
74&\includegraphics[width = 6 mm, height = 5 mm ]{id74}&$\sum{SA'\cdot\not{M}\cdot\not{M'}}$ \\
\hline
78& \includegraphics[width = 6 mm, height = 5 mm ]{id78}&$(\sum{S^2\cdot\not{M}\cdot\not{M'}}-tr{S^2})/2$\\
\hline
98&\includegraphics[width = 6 mm, height = 5 mm ]{id98}&$(\sum{A'^2\cdot{A}})/3$ \\
\hline
102&\includegraphics[width = 6 mm, height = 5 mm ]{id102}&$\sum{A^2\cdot{S}}$\\
\hline
108&\includegraphics[width = 6 mm, height = 5 mm ]{id108}&$(\sum{A'A\cdot{S}})/2$\\
\hline
110&\includegraphics[width = 6 mm, height = 5 mm ]{id110}&$\sum{S^2\cdot{A}}$\\
\hline
238&\includegraphics[width = 6 mm, height = 5 mm ]{id238}&$(\sum{S^2\cdot{S}})/6$\\
\hline
\end{tabular}\caption{Matrix formulas for the numbers of all 3-node connected directed subgraphs. $M$ is the adjacency matrix,
$S$ is its symmetric component, and $A$ its asymmetric component.
$A'$ is the transposed matrix, $\widetilde{A}$ is the logical
inverse of matrix $A$, $trA$ is the matrix trace.}\label{Table3}
\end{center}
\end{table}
We denote by $AB$ the matrix multiplication of matrices $A$ and
$B$, and by $A\cdot{B}$ the dot multiplication. $\widetilde{A}$ is
the logical inverse of matrix $A$, where the 0 elements of $A$ are
the 1 of $\widetilde{A}$ and vice-versa. $A'$ is the transpose
matrix of $A$. A summation denotes summation of all the matrix
indices. The matrix formulas for the 13 directed connected 3-node
subgraphs are given in Table \ref{Table3}. For example id38 has
two nodes which are connected by a path of 2 edges and a path of
one edge. ${A^2}_{ij}$ is the number of length 2 paths between
node i and node j. Dot-multiplication with matrix $A$ and
summation of the terms of the resultant matrix gives the correct
count. In some of the subgraphs a correction is made for the terms
on the diagonal (id6,id36,id78).\\

\begin{acknowledgments}
We thank S. Maslov, R. Cohen, A. Mayo, A. Natan, M. Itzkovitz and
all members of our lab for valuable discussions. We acknowledge
support from the Israel Science Foundation, the Human Frontier
Science Program, and the Minerva Foundation.
\end{acknowledgments}


\begin{thebibliography}{}
\bibitem{Strogatz 2001}
S. H. Strogatz, "Exploring complex networks", Nature 410, 268-76.
(2001).
\bibitem{AlbertsBarabasi 2002}
R. Albert \& A.L. Barabasi, "Statistical mechanics of complex
networks", Reviews of Modern Physics 74, 47 (2002).
\bibitem{Newman 2001}
M. Newman, S. Strogatz \& D. Watts,  "Random graphs with arbitrary
degree distribution and their applications", Phys Rev E 64,
6118-6123 (2001).
\bibitem{Dorogovtsev 2002}
S.N. Dorogovtsev, J.F.F Mendes \& A.N. Samukhin, "Principles of
Statistical Mechanics of Random Networks", cond-mat/0204111.
\bibitem{Watts 1998}
D. Watts, S. Strogatz,"Collective dynamics of 'small-world'
networks", Nature 393, 440-442 (1998).
\bibitem{Amaral2000}
L. Amaral, A. Scala, M. Barthelemy, H. Stanley,"Classes of small
world networks", Proc. Natl. Acad. Sci. U.S.A. 97, 11149-11152
(2000).
\bibitem{Shen or 2002}
S. Shen-Orr, R. Milo, S. Mangan \& U. Alon, "Network motifs in the
transcriptional regulation network of Escherichia coli", Nature
Genetics, 31:64-68 (2002).
\bibitem{Milo 2002}
R. Milo, S. Shen-Orr, S. Itzkovitz, N. Kashtan., D. Chklovskii \&
U. Alon, "Network Motifs: Simple Building Blocks of Complex
Networks", Science 298, 824-827 (2002).
\bibitem{Maslov 1998}
S. Maslov, K. Sneppen,"Specificity and Stability in Topology of
Protein Networks", Science 296, 910-3 (May 3, 2002)
\bibitem{Erdos1959}
P. Erd\H{o}s \& A. R\'{e}nyi, "On random graphs", Publicationes
Mathematicae 6, 290-297 (1959).
\bibitem{Erdos1960}
P. Erd\H{o}s \& A. R\'{e}nyi, "On the evolution of random graphs",
Publications of the Mathematical Institute of the Hungarian
Academy of Sciences 5, 17-61 (1960).
\bibitem{Erdos1961}
P. Erd\H{o}s \& A. R\'{e}nyi, "On the strength of connectedness of
a random graph", Acta Mathematica Scientia Hungary  12, 261-267
(1961).
\bibitem{Bollobas1985}
B. Bollobas, "Random Graphs", Academic Press, New York (1985).
\bibitem{Holland}
P.W. Holland, S. Leinhardt, D. Heise, "Local structure in social
networks", "Sociological Methodology" Ed. (Jossey-Bass, San
Fransisco, 1975) pp. 1-45.
\bibitem{Davis}
J.A. Davis, S. Leinhardt, "The Structure of Positive Interpersonal
Relations in Small Groups." In Joseph Berger, Morris Zelditch,
Jr., and Bo Anderson (eds.), Sociological Theories In Progress
Volume 2, Boston: Houghton Mifflin(1972), pp. 218-251.
\bibitem{Holland70}
P.W. Holland, S. Leinhardt, "A Method for Detecting Structure in
Sociometric Data" American Journal of Sociology 70(1970): 492-513
\bibitem{Wasserman}
S. Wasserman \& K. Faust, "Social Network Analysis: Methods and
Applications", Cambridge University Press, Cambridge, England,
(1994).
\bibitem{Barabasi1999}
A. L. Barabasi \& R. Albert, "Emergence of scaling in random
networks", Science 286, 509-12 (1999).
\bibitem{Redner98}
S. Redner, "How Popular is your paper? An empirical study of the
citation distribution", European Phys. J. B 4, 131 (1998).
\bibitem{Faloutsos 1999}
M. Faloutsos , P. Faloutsos , C. Faloutsos , "On power-law
relationships of the internet topology", Comp. Comm. Rev. 29,
251-262 (1999).
\bibitem{huberman}
B. A. Huberman, L. A. Adamic, "Internet: Growth dynamics of the
World-Wide Web", Nature 401, 131 (Sep 1999).
\bibitem{Burda}
Z. Burda, J.D. Correia, A. Krzywicki, "Statistical ensemble of
scale-free random graphs", Phys.Rev. E64 (2001) 046118.
\bibitem{Krzywicki}
A. Krzywicki, "Defining statistical ensembles of random graphs",
cond-mat/0110574
\bibitem{krapivsky}
P. L. Krapivsky, G. J. Rodgers, S. Redner,  "Degree Distributions
of Growing Networks",Phys. Rev. Lett. 86, 5401-5404 (2001).
\bibitem{sole}
R. Ferrer i Cancho, R.V. Sol\'{e},"The small world of human
language", Proc. R. Soc. Lond. B. 268, 2261-2266 (2001).
\bibitem{valverde}
S. Valverde, R. Ferrer i Cancho, R. V. Sole,"Scale-free Networks
from Optimal Design", cond-mat/0204344 (2002)
\bibitem{Dorogovtsev_evolution}
S.N. Dorogovtsev, J.F.F. Mendes,, "Evolution of networks", Adv.
Phys. 51, 1079-1187 (2002)
\bibitem{dorogovtsev_mesoscopic}
S.N. Dorogovtsev, A.N. Samukhin, "Mesoscopics and fluctuations in
networks", Phys. Rev. E 67, 037103 (2001)
\bibitem{Sole 2001}
R. Ferrer i Cancho, C. Janssen \& R. V. Sole, , "Topology of
technology graphs: small world patterns in electronic circuits".
Physical Review E, 64, 046119, (2001).
\bibitem{Aiello 2001}
W. Aiello, F. Chung \& L. Lu, "A random graph model for power law
graphs", Experiment. Math. 10 (2001), 53-66.
\bibitem{Molloy 1998}
M. Molloy \& B. Reed, "The size of the giant component of a random
graph with a given degree sequence" , Combinatorics, Probability
and Computing 7, 295-305 (1998).
\bibitem{Molloy 1995}
M. Molloy \& B. Reed, "A critical point for random graphs with a
given degree sequence" , Random Structures and Algorithms 6,
161-179 (1995).
\bibitem{Bender}
E. Bender, E. Canfield, "The asymptotic number of labelled graphs
with given degree sequences", J. Combin. Theory Ser. A 24, 296-307
(1978).
\bibitem{Chung_diameter}
F. Chung, L. Lu, "The average distances in random graphs with
given expected degrees", Proc. Natl. Acad. Sci. U.S.A., 99,
15879-15882 (2002).
\bibitem{Cohen 2002}
R. Cohen, D. Ben-Avraham, S. Havlin, "Percolation Critical
Exponents in Scale-Free Networks",Phys. Rev. E 66, 036113 (2002)
\bibitem{Cohen_resilience}
R. Cohen, K. Erez, D. ben-Avraham, S. Havlin, "Resilience of the
Internet to Random Breakdowns", Phys. Rev. Lett. 85, 4626 (2000)
\bibitem{Newman 1999 percolation}
M. Newman, "Scaling and percolation in the small-world network
model",Phys. Rev. E 60, 7332-7342 (1999)
\bibitem{Newman book}
M. Newman, "Random graphs as models of networks", Handbook of
Graphs and Networks, edited by S. BornHoldt \& G. Schuster
(Wiley-VCH, Berlin 2002).
\bibitem{Eckmann}
J. Eckmann, E. Moses, "Curvature of co-links uncovers hidden
thematic layers in the World Wide Web", Proc. Natl. Acad. Sci.
U.S.A. 99, 5825-5829 (2002)
\bibitem{Eckmann2}
P. Collet, J. Eckmann, "The Number of Large Graphs with a Positive
Density of Triangles", Journal of Statistical Physics, 2002, Vol.
108, n. 5-6, p. 1107-24.
\bibitem{dorogovtsev 2002a}
S.N. Dorogovtsev, J.F.F Mendes \& A.N. Samukhin, "Modern
architecture of random graphs: Constructions and correlations ",
cond-mat/0206467.
\bibitem{Ravasz 2002}
E. Ravasz, A. L. Barabasi, "Hierarchical Organization in Complex
Networks", Physical Review E (in press).
\bibitem{Maslov B}
S. Maslov, K. Sneppen, A. Zaliznyak, "Pattern Detection in Complex
Networks: Correlation Profile of the Internet",
cond-mat/0205379(2002)
\bibitem{Ouzonis 2000}
C.A. Ouzounis, P.D. Karp, "Global properties of the metabolic map
of Escherichia coli", Genome Research 10, 568-576  (2000)
\bibitem{Wagner 2002}
A. Wagner, D. Fell, "The small world inside large metabolic
networks", Proc. R. Soc. Lond. B. 2001 Sep 7;268(1478):1803-10.
\bibitem{Guelzim 2002}
N. Guelzim, S. Bottani, P. Bourgine, F. K\'{e}p\`{e}s,
"Topological and causal structure of the yeast transcriptional
regulatory network", Nature Genet. 31, 60(2002).
\bibitem{bianconi_hloops}
G. Bianconi, A. Capocci, "Number of Loops of Size h in Growing
Scale-Free Networks" Phys. Rev. Lett. 90, 078701 (2003).
\bibitem{berg}
J. Berg , M. L\"{a}ssig, "Correlated Random Networks", Phys. Rev.
Lett. 89 (22),228701 (2002)
\bibitem{Newman 1998}
M. Newman, "Assortative mixing in networks", Phys. Rev. Lett. 89,
208701 (2002).
\bibitem{Bianconi 2001}
G. Bianconi, A.L. Barabasi, "Bose-Einstein condensation in complex
networks" Physical Review Letters 86, 5632-5635 (2001). P.
\bibitem{White}
J.G. White, E. Southgate, J.N. Thomson \& S. Brenner, "The
structure of the nervous system of the nematode Caenorhabditis
elegans", Phil. Trans. Roy. Soc. London Ser. B, 314, 1-340 (1986).
\bibitem{Harari}
F. Harary, H.J. Kommel," Matrix measures for transitivity and
balance", Journal of Mathematical Sociology, Vol 6.: 199-210
(1979).
\bibitem{Footnote1}
In subgraphs which have several nodes with the maximal degree,
this approximation should still give correct scaling results as
long as the probability of obtaining several large hubs in one
subgraph is low. In obtaining the scaling relations, we replace
the number of mutual edges with their average value
$\mean{K}^2/N$. This gives an incorrect value for the exact number
of appearances (compare id14 vs. id38 in Table 1), but is valid
for obtaining scaling relations.
\end{thebibliography}
\end{document}